\documentclass[sigconf]{acmart}
\AtBeginDocument{%
  }

\setcopyright{acmlicensed}
\copyrightyear{2018}
\acmYear{2018}
\acmDOI{XXXXXXX.XXXXXXX}
\acmConference[Conference acronym 'XX]{Make sure to enter the correct
  conference title from your rights confirmation email}{June 03--05,
  2018}{Woodstock, NY}
\acmISBN{978-1-4503-XXXX-X/2018/06}

\begin{document}

\title{Structured Multi-Step Reasoning for Entity Matching \\Using Large Language Model}

\author{Rohan Bopardikar}
\email{rbopardi@asu.edu}
\affiliation{%
  \institution{Arizona State University}
  \city{Tempe}
  \state{Arizona}
  \country{USA}
}

\author{Jin Wang}
\email{jinwang18@asu.edu}
\affiliation{%
  \institution{Arizona State University}
  \city{Tempe}
  \state{Arizona}
  \country{USA}
}

\author{Jia Zou}
\email{jia.zou@asu.edu}
\affiliation{%
  \institution{Arizona State University}
  \city{Tempe}
  \state{Arizona}
  \country{USA}
}

\renewcommand{\shortauthors}{Bopardikar et al.}

\begin{abstract}
Entity matching is a fundamental task in data cleaning and data integration. With the rapid adoption of large language models (LLMs), recent studies have explored zero-shot and few-shot prompting to improve entity matching accuracy. However, most existing approaches rely on single-step prompting and offer limited investigation into structured reasoning strategies. In this work, we investigate how to enhance LLM-based entity matching by decomposing the matching process into multiple explicit reasoning stages. We propose a three-step framework that first identifies matched and unmatched tokens between two records, then determines the attributes most influential to the matching decision, and finally predicts whether the records refer to the same real-world entity. In addition, we explore a debate-based strategy that contrasts supporting and opposing arguments to improve decision robustness. We evaluate our approaches against multiple existing baselines on several real-world entity matching benchmark datasets. Experimental results demonstrate that structured multi-step reasoning can improve matching performance in several cases, while also highlighting remaining challenges and opportunities for further refinement of reasoning-guided LLM approaches.
\end{abstract}


\maketitle
\pagestyle{plain}

\section{Introduction}\label{sec-intro}

Entity Matching (EM), also known as Entity Resolution or Record Linkage, refers to the problem of deciding whether a given pair of entity records refers to the same object in real world~\cite{li2021deep}.
It is a fundamental problem in data integration and has a wide spectrum of real world applications, such as data cleaning, knowledge base construction, clustering and web search.

An end-to-end EM pipeline consists of three stages~\cite{jedai2019,magellan2016}: 
\emph{Blocking} takes two tables of raw entity records as input and returns a set of candidate pairs that might be matched; 
\emph{labeling} aims at identifying a small subset from the candidate set and generates the training and validation sets; 
\emph{matching} utilizes such training data to train a high-quality matcher model to be deployed for the prediction of matching result.
In this paper, we focus on the works in the \emph{matching stage}.
There is a long stream of studies to improve the matcher model in the database community.
Earlier studies aimed at developing rule based method or employing classic machine learning techniques to build a binary classifier to make predictions~\cite{elmagarmid2006duplicate,kopcke2010frameworks}.
In the past decade, deep learning techniques have been widely adopted in the EM problem and achieved very promising results~\cite{sigmod2018deepem,joty2018distributed}.
Following this route, recent studies regarded EM as a sequential pair classification task and fine-tunes the pre-trained language model (PLM) such as BERT to build the matcher model~\cite{pvl2020bert,transformer_em2020,wang2022machop}.
Such PLM based solutions showed significant improvement and achieved state-of-the-art performance.

Most recently, advances in the era of Large Language Models (LLMs) have brought new opportunities to the EM problem. 
Pre-trained LLMs such as GPT, Gemini and Llama have shown superior abilities in understanding human instructions as well as generating structured output. 
It has been well known that the input instructions to LLM, which is known as \emph{prompt}, is essential to achieve desired output.
As a result, the LLM-based solutions focused on effective \emph{prompt engineering} strategies to improve the overall performance.
And there have been some efforts in this field for the EM task~\cite{llm_em_edbt2025,chatgpt_em,costeffective2024,narayan2022can}.

Nevertheless, there is still room to further improve the LLM based methods for Entity Matching.
While previous efforts aimed at developing effective prompt templates to provide meaningful instructions as well as rich contextual information to guide the LLMs, little attention has been paid in taking advantage of the powerful reasoning capability of LLMs.
To fill this gap, in this paper we make some preliminary investigations about how to employ reasoning strategies to improve the LLM based prompt engineering for Entity Matching.
We recognize from the series of studies about the explainable entity matching~\cite{wang2022minun,teofili2022effective,benassi2024explaining,shahbazi2024fairness} that the matching between the pair of entitles could be based on matching between attributes and tokens from them, respectively. 
This findings perfectly aligns with the high level idea of the Chain-of-Thought (CoT) reasoning~\cite{wei2022chain} strategy.
Based on this observation, we developed a 3-step reasoning strategy that asks LLM to first look for the matching information in the token level and attribute level, and finally make the prediction of the entity-level matching accordingly.
We implemented this idea with two methods and evaluated on 6 popular benchmarking tasks.
The initial results illustrated some benefit in the improvement of effectivness.

The rest of this paper is organized as following: Section~\ref{sec-related} surveyed the related work; Section~\ref{sec-prelim} introduced some background knowledge; Section~\ref{sec-method} introduced the proposed methods; Section~\ref{sec-exp} reported the initial experimental results; finally Section~\ref{sec-conc} concluded the whole paper.

\section{Related Works}\label{sec-related}

\noindent\textbf{End-to-End Entity Matching Systems}
Early research on entity matching emphasized building comprehensive systems that support the entire matching pipeline, from data pre-processing and blocking to matching and evaluation. Magellan~\cite{magellan2016} is one of the earliest entity matching management systems that provides modular capabilities for feature engineering, blocking, and supervised learning, establishing a foundation for systematic entity matching workflows. JedAI~\cite{jedai2019} further advanced this direction by offering a domain- and structure-agnostic framework for scalable, end-to-end entity resolution with automated workflow construction. Konda et al~\cite{endtoend2019} demonstrated practical deployment of complete matching pipelines and highlighted the importance of orchestration and system integration in real-world scenarios. These works primarily focus on pipeline execution and system design, rather than the reasoning mechanisms underlying entity matching decisions.

\noindent\textbf{Deep Learning-based Entity Matching}
The adoption of deep learning significantly transformed entity matching by enabling automated representation learning. Early work explored the use of recurrent neural networks (RNNs) and conducted extensive design space analyses for deep matching architectures~\cite{sigmod2018deepem}. Later, transformer architectures and PLMs such as BERT brought substantial improvements by capturing contextual semantics in textual attributes~\cite{pvl2020bert,transformer_em2020}.
Recent studies extended deep learning to more challenging settings. Sudowoodo introduced contrastive self-supervised learning for multi-purpose data integration, improving generalization across tasks~\cite{sudowoodo2023}. Ground truth inference for weakly supervised entity matching explored methods to reduce reliance on labeled data~\cite{weaklysigmod2023}. Despite their effectiveness, these approaches largely treat entity matching as a single-step classification problem without explicitly modeling intermediate reasoning or attribute-level contribution.
\smallskip

\noindent\textbf{LLM-based Entity Matching}
With the emergence of large language models (LLMs), several works have explored prompt-based entity matching in zero-shot and few-shot settings. Studies investigated different prompting strategies such as matching, comparing, or selecting records~\cite{matchcompareselect2025} and evaluated the feasibility of LLMs for direct entity matching~\cite{llm_em_edbt2025}. Other efforts examined cost-effective in-context learning and design trade-offs to balance performance and computational efficiency~\cite{costeffective2024}.
Some works further explored fine-tuning LLMs for entity matching, showing potential performance gains compared to purely prompt-based approaches~\cite{finetune_llm_em}. Similarly, prior studies assessed the applicability of ChatGPT and general-purpose LLMs for entity matching~\cite{chatgpt_em}. However, most LLM-based methods rely on single prompt and lack structured, interpretable multi-step reasoning. Our work differs by explicitly decomposing the matching process into token comparison, attribute importance analysis, and decision making, enabling more transparent and accurate entity matching.
\smallskip

\noindent\textbf{Experimental Studies, Benchmarks, and Datasets}
A separate line of work focuses on systematic evaluation and benchmarking of entity matching methods. Progressive Entity Matching provides a broad design space exploration across multiple matching strategies~\cite{progressive2025}. Large-scale datasets such as WDC Products introduce multi-dimensional benchmarks that reflect real-world e-commerce scenarios~\cite{wdc2024}. 
Other studies investigated individual components of the pipeline, such as blocking and filtering techniques. Sparkly demonstrated the strength of TF-IDF-based blocking methods~\cite{sparkly2023}, while comparative studies analyzed approximate blocking strategies and filtering techniques~\cite{benchmarkblocking2023,blocking2016}. Additional works evaluated the performance of PLMs and embeddings for entity resolution tasks~\cite{bertanalysis2022,embedding2023}. These studies provide critical insights into evaluation methodology and dataset diversity, which we adopt in our experimental design.
\smallskip

In summary, prior work spans system-level platforms~\cite{magellan2016,jedai2019,endtoend2019}, deep learning and PLM-based approaches~\cite{sigmod2018deepem,pvl2020bert,transformer_em2020,sudowoodo2023,weaklysigmod2023}, and prompt-based LLM methods~\cite{matchcompareselect2025,llm_em_edbt2025,costeffective2024,finetune_llm_em,chatgpt_em}. While these approaches have advanced entity matching accuracy and scalability, most treat matching as a monolithic prediction task. In contrast, our work introduces a structured multi-step reasoning framework that explicitly models intermediate evidence, offering improved accuracy and interpretability for LLM-based entity matching.

\section{Preliminaries}\label{sec-prelim}

\subsection{Problem Statement.}

Entity Matching (EM), also known as entity resolution or record linkage, aims to identify records that refer to the same real-world entity across one or more data sources~\cite{magellan2016,li2021deep}. Each object (or record) $o \in \mathcal{O}$ is represented as a tuple of attributes $o = \langle a_1, a_2, \dots, a_m \rangle$, where each attribute may contain different types of values.

Since exhaustive comparison of all record pairs is computationally prohibitive, existing EM pipelines first apply a \emph{blocking} or \emph{filtering} strategy to group potentially matching records into blocks of candidate pairs~\cite{blocking2016,sparkly2023}. Formally, given a dataset $\mathcal{O}$, a blocking function partitions objects into a set of blocks $\{\mathcal{B}_1, \mathcal{B}_2, \dots, \mathcal{B}_k\}$, where only record pairs within the same block are considered for matching.

For each candidate pair $\langle o_1, o_2 \rangle \in \mathcal{B}_i \times \mathcal{B}_i$, the matching decision is defined as a binary classification task:
\[
match(o_1, o_2) \in \{0,1\},
\]
where $match(o_1, o_2) = 1$ indicates that $o_1$ and $o_2$ refer to the same real-world entity, and $0$ otherwise.

In this work, we instead focus on an LLM-based reasoning formulation. Specifically, we define an LLM-based matching function:
\[
\mathcal{F}_{\text{LLM}} : (o_1, o_2) \rightarrow \hat{t}, \quad \hat{t} \in \{0,1\},
\]
where $\hat{t}$ denotes the predicted match label.

Given the ground-truth label $t \in \{0,1\}$, our objective is to design a structured reasoning strategy for $\mathcal{f}_{\text{LLM}}$ that maximizes the probability of producing the correct matching decision:
\[
\max_{\mathcal{F}_{\text{LLM}}} \; P(\hat{t} = t \mid o_1, o_2).
\]

Unlike prior methods that treat matching as a single-step prediction task, our formulation explicitly decomposes $\mathcal{F}_{\text{LLM}}$ into intermediate reasoning stages that guide the LLM through token-level comparison and attribute-aware analysis before reaching the final decision. This design aims to improve both predictive accuracy and interpretability of LLM-based entity matching.

\subsection{LLM and Reasoning Strategies}
Large Language Models (LLMs) demonstrate strong capabilities in performing complex reasoning tasks through prompt-based interaction, enabling them to move beyond surface-level pattern matching toward structured decision making. Among various reasoning paradigms, \emph{chain-of-thought} (CoT) reasoning~\cite{wei2022chain} encourages the model to articulate intermediate reasoning steps that lead to a final answer, thereby improving transparency and logical consistency, while \emph{debate-based reasoning} prompts the model (or multiple agents) to construct and reconcile opposing arguments before producing a final judgment, leading to more balanced and robust decisions~\cite{du2023improving, liang2024encouraging}. Both strategies have been shown to significantly enhance performance on tasks requiring logical inference and contextual understanding.

In this work, we focus on these two reasoning methodologies because they naturally align with the inherent structure of the entity matching problem. Entity matching requires both fine-grained analytical reasoning (e.g., comparing token-level similarities and assessing attribute importance) and critical evaluation of ambiguous evidence. Chain-of-thought reasoning enables step-by-step decomposition of the matching process, making intermediate evidence explicit and interpretable, while debate-based reasoning encourages the model to examine conflicting signals and mitigate overconfident or shallow decisions. Together, these strategies provide complementary strengths: CoT improves structured evidence extraction, and debate-based reasoning strengthens decision robustness, making them particularly suitable for reliable and interpretable LLM-based entity matching.

\subsection{Existing General Prompt Template of LLM-based Entity Matching}\label{subsec-baseline}

\begin{figure}[h]
\centering{%
   \includegraphics[width=3.3in]{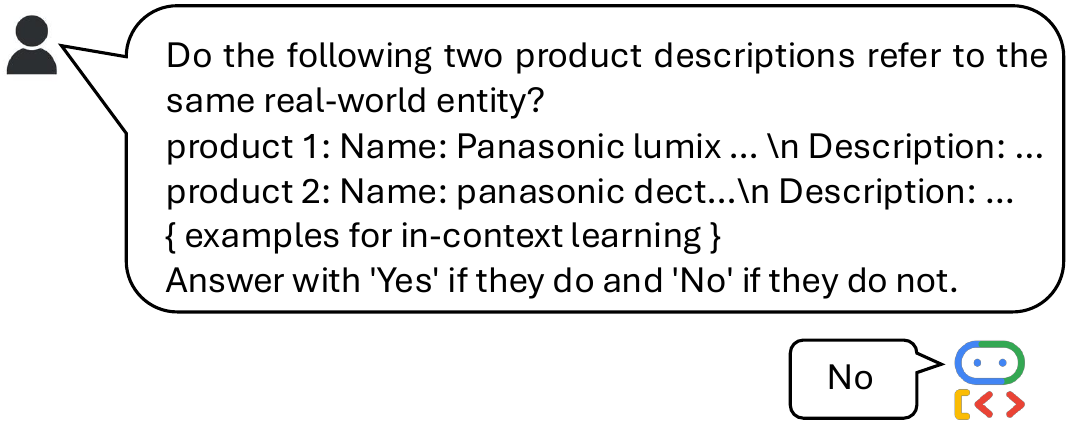}  
}
\caption{\label{fig:general-example} \small {Illustration of a general prompt for entity matching}}
\end{figure}

In the context of entity matching, most existing LLM-based approaches rely on single-step prompting that directly produces a binary match decision~\cite{llm_em_edbt2025}.
This might limit the transparency and fail to leverage rich intermediate signals such as token-level similarity and attribute-level importance. An existing general prompt template is illustrated in Fig.~\ref{fig:general-example}.

\section{Methodology}\label{sec-method}

To fill this gap, we view LLM reasoning as a structured, staged process that progressively refines the matching decision through explicit intermediate analysis. This perspective aligns with emerging reasoning-aware paradigms, including structured multi-step reasoning and debate-based deliberation, which have been shown to improve both decision accuracy and interpretability in LLM-driven tasks~\cite{zhu2025perception}. In this section, we will first explore the design space of general entity-matching prompting strategies and then propose two multi-step reasoning frameworks for the entity-matching tasks.

\subsection{Design Space Exploration}
We first explore the design space of a general prompting strategy (single-step or multi-step) from the following dimensions.

\noindent
\textbf{Task Frame: General or Domain-Specific.}
We investigate whether the task instruction should remain general (e.g., ``Determine if these two objects match'') or incorporate domain-specific descriptions (e.g., ``Determine if these two product listings represent the same consumer electronic device''). Domain-aware prompts may improve performance by supplying implicit semantic priors but risk reducing generality.

\noindent
\textbf{Verbiage Complexity: Simple or Complex.} We compare prompts with concise, direct instructions against more descriptive, explanatory prompts. This analysis examines the trade-off between clarity and cognitive load imposed on the LLM.

\noindent
\textbf{Response Frame: Free or Forced.} We evaluate whether allowing open-ended responses or constraining outputs to predefined templates (e.g., \texttt{Match: Yes/No}) better supports reliable decision making. Forced-response formats improve consistency and simplify parsing, while free-text responses offer richer interpretability.

\noindent
\textbf{In-Context Learning.} We explore the impact of providing exemplar matching cases in the prompt, including positive and negative examples, to examine how few-shot learning influences reasoning accuracy and generalization.

\noindent
\textbf{Additional hints: Exact Matched Words or Phrases.}
We test the effectiveness of explicitly injecting matched token highlights or pre-identified common phrases into the prompts. These hints can guide attention toward key evidence but may introduce bias if poorly selected.

\subsection{A Three-Step Reasoning Framework.} \label{subsec-reasoning}
We first propose a three-step reasoning strategy, where the input of current step is the output of a previous step, as follows: 

\begin{itemize}
\item\textbf{Step 1.} Ask LLM to identify matched and unmatched tokens between the two objects;
\item\textbf{Step 2.} Ask LLM to identify which attributes are most influential for making the match decision;
\item \textbf{Step 3.} Based on steps 1 and 2, ask LLM whether these objects refer to the same real-world entity.
\end{itemize}

\begin{figure}[t]
\centering{%
   \includegraphics[width=3.3in]{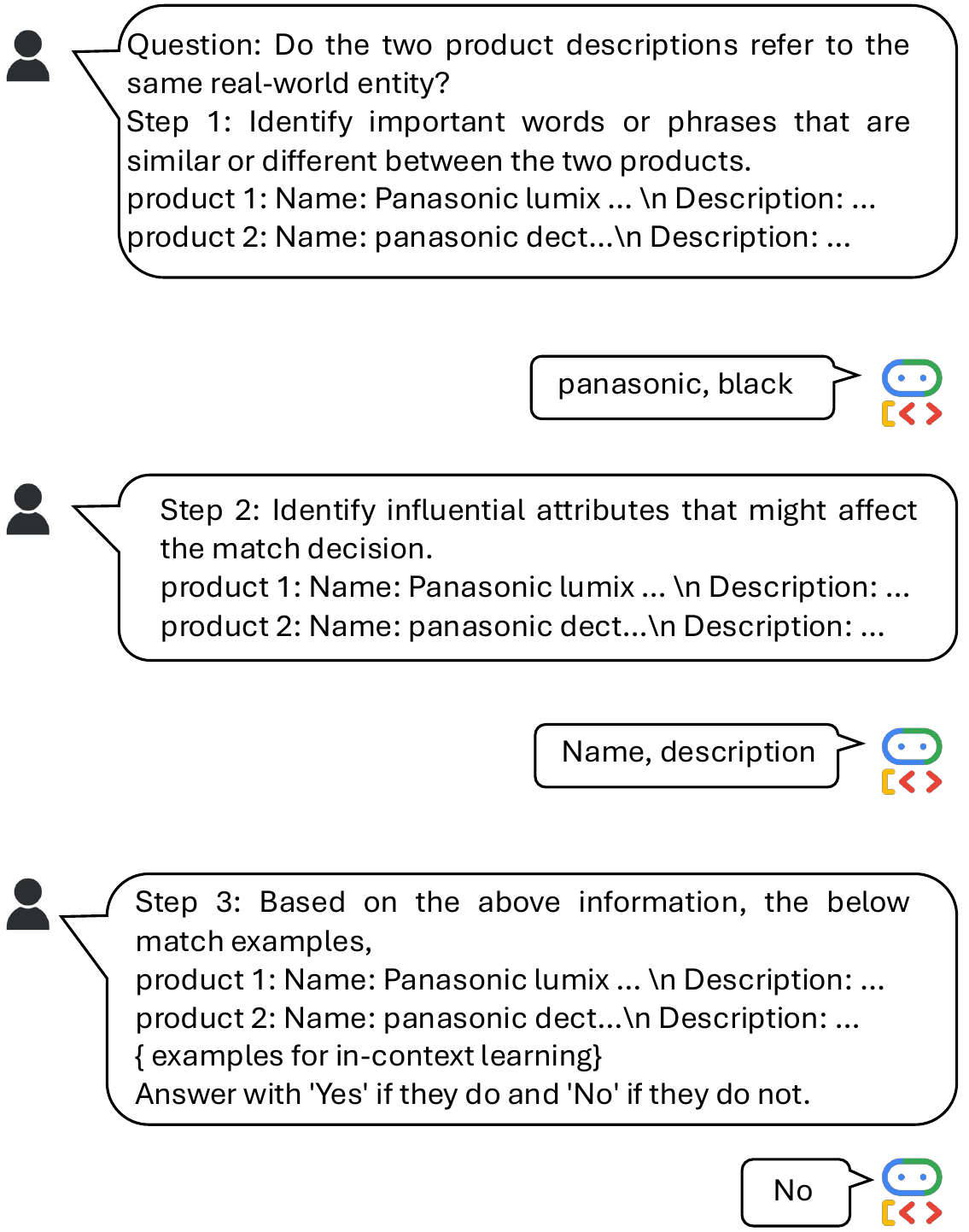}  
}
\caption{\label{fig:multi-step-multi-prompt} \small {Three-step reasoning with multiple prompts}}
\end{figure}

\begin{figure}[h]
\centering{%
   \includegraphics[width=3.3in]{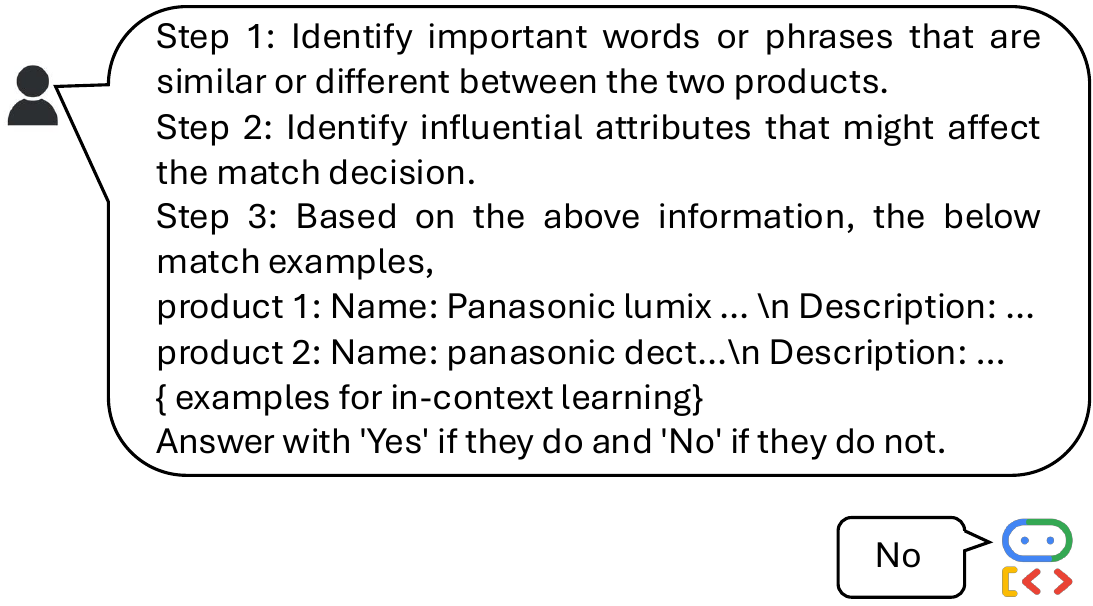}  
}
\caption{\label{fig:multi-step-single-prompt} \small {Three-step reasoning with a single prompt}}
\end{figure}

We have two configurations for this framework: \textit{1. Multi-Prompts}, where each step is a sequential chaining of multiple prompts, with each prompt corresponding to one step, as illustrated in Fig.~\ref{fig:multi-step-multi-prompt}. The LLM response of the previous step will be appended to the next-step prompt.  \textit{2. Single-Prompt}, where all three steps are encapsulated in one prompt, as illustrated in Fig.~\ref{fig:multi-step-single-prompt}.

\subsection{A Debate-based Reasoning Framework.} \label{ubsec-deb}
This design is inspired by prior work on debate-based and multi-agent reasoning strategies~\cite{du2023improving, liang2024encouraging}. The intuition behind the debate-based strategy is to explicitly surface both supporting and opposing evidence before making a final matching decision. Rather than directly producing a binary judgment, the LLM is first guided to construct structured arguments for both sides of the decision, thereby encouraging more balanced and self-critical reasoning.

\begin{figure}[h]
\centering{%
   \includegraphics[width=3.3in]{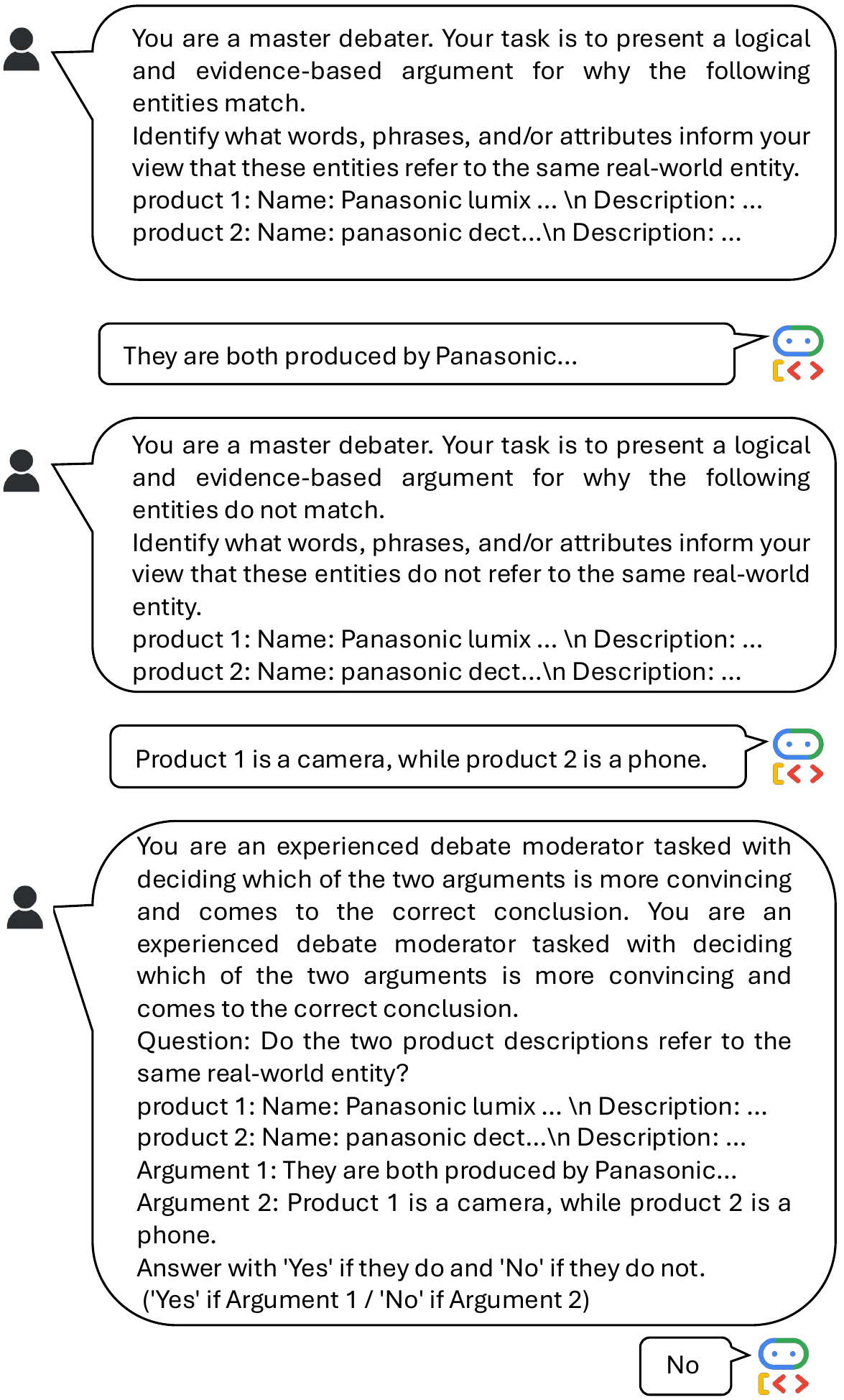}  
}
\caption{\label{fig:debate-prompt} \small {Debate-based reasoning}}
\end{figure}

As illustrated in Fig.~\ref{fig:debate-prompt}, we prompt the LLM to generate two complementary perspectives: 
(i) reasons why the given pair of objects \emph{should} refer to the same real-world entity, and (ii) reasons why they \emph{should not} match. These pro and con arguments are then provided back to the LLM in a subsequent step, where it is asked to synthesize both viewpoints and produce a final, justified decision. This process encourages the model to compare competing evidence, reduce overconfidence, and mitigate shallow pattern-based judgments. Each step is executed using a separate prompt.

\section{Experimental Evaluation}\label{sec-exp}

\subsection{Experiment Setup}
\begin{table}[ht]\scriptsize
	\centering
	\caption{The statistics of datasets}\label{tbl-stats}
    \vspace{-1em}
	\begin{tabular}{l|cccc}
		\toprule
		Dataset & Domain & \# Positive & \# Negative  & \# Attributes \\
		\midrule
		Abt-Buy (AB) & product & 1,028 & 8,547 & 3  \\
		DBLP-ACM (DA) & citation & 2,220 & 10,143 & 4 \\
        DBLP-Scholar (DS) & citation & 5,347 & 23,360  & 4 \\
        Walmart-Amazon (WA) & electronics & 962 & 9,280 &  5\\
        Amazon-Google (AG) & software & 1,167 & 10,293 &  3\\
        WDC & product & 2,250 & 7,992 &  3 \\
		\bottomrule
	\end{tabular}
\end{table}

\noindent\textbf{Datasets}\hspace{.5em}
We conduct experiments on 6 public benchmarking tasks that are widely evaluated in previous studies. 
The first 5 datasets are provided by the DeepMatcher~\cite{deepmatcher_datasets} project.
These datasets are for training and evaluating EM models for various domains including products, publications, and businesses. 
Each dataset provides sets of labeled entity pairs where each pair has a binary label indicating whether it is a match or not.
The WDC dataset~\cite{wdc2024} is about the products matching about records from thousands of e-shops.
We use the hardest version that includes 80\% corner cases.
The detailed information is shown in Table~\ref{tbl-stats}.
\smallskip

\noindent\textbf{Evaluation Metrics}\hspace{.5em}
In this experiment, we report the results of both effectiveness and LLM cost.
For effectiveness, we use $F_1$ score as the primary metric following the previous studies.
For LLM coat, we report the total number of input and output tokens for each method since the pricing of OpenAI APIs is based on the usage of tokens.
\smallskip

\noindent\textbf{Environment}\hspace{.5em} 
All experiments were run on a PowerEdge R750 server with 24 cores, 128 GB memory and 2TB HDD.
For the prompts over LLM, we use the official OpenAI APIs of GPT-5.1-mini and obtain the response as the final results.

\subsection{Preliminary Results}

\begin{table}[ht]
	\centering
	\caption{Results for Zero-shot Learning.}\label{tbl-zeros}
    \vspace{-1em}
	\begin{tabular}{l|cc|cc|cc}
		\toprule
		Method & \multicolumn{2}{c|}{Baseline} & \multicolumn{2}{c|}{Single Prompt} & \multicolumn{2}{c}{Multi-Prompt} \\
		& F1 & \# Tokens  & F1 & \# Tokens & F1 & \# Tokens \\
		\midrule
		AB & 0.849 & 183 & 0.849 & 559 & \textbf{0.862} & 1499 \\
        DA & \textbf{0.949} & 146 & 0.929 & 506 & 0.948 & 1274 \\
        DS & 0.899 & 133 & 0.899 & 499 & \textbf{0.918} & 1332 \\
        WA & \textbf{0.79} & 149 & 0.743 & 519 & 0.756 & 1508 \\
        AG & 0.581 & 104 & \textbf{0.628} & 459 & 0.652 & 1245 \\
        WDC & 0.798 & 326 & \textbf{0.831} & 695 & 0.815 & 2124 \\
		\bottomrule
	\end{tabular}
\end{table}

Here we report the results of three methods: \textsf{Baseline} means the general prompt strategy without reasoning method introduced in Section~\ref{subsec-baseline};
\textsf{Single-Prompt} and \textsf{Multi-Prompt} is the method that includes the reasoning steps within one prompt and into different prompts introduced in Section~\ref{subsec-reasoning}, respectively.
For the Debate based strategy introduced in Section~\ref{ubsec-deb}, we recognize that it is significantly worse than the other three methods. 
For example, on the DA dataset, the debate based prompt only achieves 0.835 in $F_1$ score under zero-shot setting.
Thus we believe it is not a feasible strategy for entity matching and omit the results from comparison.
We report the results with both zero-shot and few-shot learning settings.

The results of zero-shot learning are shown in Table~\ref{tbl-zeros}. 
We can see that reasoning based methods achieve better performance in 4 out of 6 tasks.
This illustrates that the reasoning strategies could bring potential benefits to the entity matching task.
At the same time, \textsf{Multi-Prompt} has a slightly better overall performance compared with \textsf{Single-Prompt}.
However, the LLM cost is much heavier since each step requires seperate input and output tokens.

The results of few-shot learning are shown in Table~\ref{tbl-fews}.
Here we choose the number of examples as 2 empirically.
We observe that the relative performance of reasoning based methods are not as good as that under zero-shot settings.
The reason might be that the LLM could obtain some useful signals from the examples and therefore the insights brought by the reasoning strategy become less important.

To sum up, the reasoning based methods have showed certain performance gain even with very naive implementations.
Thus it is promising to continue investigating more advanced strategies following this route.
However, its heavy LLM cost is also a potential issue to be resolved so as to make a good tradeoff between effectiveness and overhead.

\begin{table}[ht]
	\centering
	\caption{Results for 2-shot Learning.}\label{tbl-fews}
    \vspace{-1em}
	\begin{tabular}{l|cc|cc|cc}
		\toprule
		Method & \multicolumn{2}{c|}{Baseline} & \multicolumn{2}{c|}{Single Prompt} & \multicolumn{2}{c}{Multi-Prompt} \\
		& F1 & \# Tokens  & F1 & \# Tokens & F1 & \# Tokens \\
		\midrule
		AB & \textbf{0.9} & 485 & 0.895 & 868 & 0.878 & 1897 \\
        DA & \textbf{0.978} & 371 & 0.94 & 729 & 0.956 & 1496 \\
        DS & 0.899 & 333 & 0.91 & 697 & \textbf{0.922} & 1333\\
        WA & \textbf{0.816} & 387 & 0.797 & 757 & 0.771 & 1748 \\
        AG & 0.678 & 249 & \textbf{0.721} & 604 & 0.672 & 1388 \\
        WDC & 0.818 & 798 & \textbf{0.833} & 1183 & 0.822 & 2154 \\
		\bottomrule
	\end{tabular}
\end{table}

\section{Conclusions}\label{sec-conc}

In this paper, we made some initial investigation the utilize reasoning strategies to improve the performance of LLM based solution for Entity Matching.
Based on the high level idea of chain-of-thought reasoning, we split the process of entity matching into three steps and guide the LLMs to make the final prediction in a step-by-step manner. 
We implemented this idea in two methods with one prompt and multiple prompts, respectively.
We conducted the experiments on six public benchmarking tasks.
The initial results showed certain benefits of reasoning strategies but also indicated some rooms for further improvements in the future work.

\balance
\bibliographystyle{ACM-Reference-Format}
\bibliography{refs}
\end{document}